\documentclass[graybox]{svmult}
\usepackage{hyperref}
\usepackage{academicons}
\usepackage{xcolor}
\usepackage{multirow}
\usepackage{helvet}         
\usepackage{courier}        
\usepackage{type1cm}        
\usepackage{amsmath}
\usepackage{graphicx}        
                             \usepackage{listing}
\graphicspath{{Figures/}}
\usepackage{multicol}        
\usepackage[bottom]{footmisc}

\begin{document}

\title{Cloud Big Data Mining and Analytics: Bringing Greenness and Acceleration in the Cloud}
\author{Hrishav Bakul Barua and Kartick Chandra Mondal}
\institute{Hrishav Bakul Barua \at Embedded Systems and Robotics Research Group (Cognitive Robotics), TCS Research, Kolkata \email{hrishav.barua@tcs.com, hbarua@acm.org} 
\and Kartick Chandra Mondal \at Department of Information Technology, Jadavpur University, Saltlake, Kolkata - 106 \email{kartick.mondal@jadavpuruniversity.in}
}

\maketitle

\abstract{
    Big data is gaining overwhelming attention since the last decade. Almost all the fields of science and technology have experienced a considerable impact from it.
    The cloud computing paradigm has been targeted for big data processing and mining in a more efficient manner using the plethora of resources available from computing nodes to efficient storage.
    Cloud data mining introduces the concept of performing data mining and analytics of huge data in the cloud availing the cloud resources.
    But can we do better? Yes, of course!
    The main contribution of this chapter is the identification of four game-changing technologies for the acceleration of computing and analysis of data mining tasks in the cloud.
    Graphics Processing Units can be used to further accelerate the mining or analytic process, which is called GPU accelerated analytics.
    Further, Approximate Computing can also be introduced in big data analytics for bringing efficacy in the process by reducing time and energy and hence facilitating greenness in the entire computing process.
    Quantum Computing is a paradigm that is gaining pace in recent times which can also facilitate efficient and fast big data analytics in very little time.
    We have surveyed these three technologies and established their importance in big data mining with a holistic architecture by combining these three game-changers with the perspective of big data.
    We have also talked about another future technology, i.e., Neural Processing Units or Neural accelerators for researchers to explore the possibilities.
    A brief explanation of big data and cloud data mining concepts are also presented here.
}

\section{Introduction}
\label{sec:introduction}

Classical data mining algorithms have been very successful in almost all the fields of science and technology ranging from medical analytics to space sciences.
The classical concepts of data mining are centered around four main paradigms: Descriptive (Clustering), Associative (Associative Rule Mining), Discriminant (Classification), and Predictive analysis (Regression).
However, with the growing amount of data, it is becoming extremely difficult to manage and process it. Retrieving information is also becoming a headache.
So, people are moving towards distributed and parallel computing paradigms \cite{barua2019comprehensive}.
But, using such paradigms in a standalone system can be another challenge in terms of scalability and cost-effectiveness, so the concept of data mining in the cloud has come up in the last decade \cite{barua2019comprehensive}.
The paper \cite{barua2019comprehensive} has summarized the recent applications, trends, techniques, algorithms, and frameworks in cloud data mining and big data mining in the cloud.

Now, is this enough? Certainly not!
The way in which data is exponentially increasing, simply harnessing the sea of resources (compute, storage, etc.) available in the cloud even may not be just enough.
So, researchers are keen in finding alternatives and effective solutions to this data outburst or we can say "data apocalypse".
Graphics Processing Units (GPUs) are the most important hardware assets in this respect.
GPUs are a special kind of CPUs which are designed for parallel computing and specifically used to alter memory in a very rapid manner.
They are designed for processing graphics related calculations in an efficient fashion.
GPUs are being used for various data mining and machine learning tasks for accelerating the entire process of analytics and mining \cite{bhargavi2017accelerating}.

Approximate Computing (AC) or In-exact Computing is a computing paradigm that trades off energy in return of accuracy of results.
Some applications which involve visual data processing, audio signal processing, big data mining and such other applications do not need exact computational results most of the time. A bit of deviation in results may always be accepted and awarded in terms of time and energy savings.
Some works are well in place which gives the idea of AC and its use in big data analytics and mining related tasks \cite{nair2014big,barua2019approximate}.

Quantum Computing (QC) is a buzz word in the today's computing community.
The concept of QC is based on the superposition of states of a computer (other than 0 or 1).
The quantum computers can solve many computing intensive problems in a much lesser time than their classical counterparts.
Data mining and machine learning techniques are well set to utilize this paradigm as and when it grows and keeps on maturing with time \cite{wittek2014quantum}.
Its popularity has grown to such an extent that a journal named "Quantum machine intelligence" (on quantum artificial intelligence) \cite{acampora2019quantum} has been started by Springer Nature recently.

\begin{figure}[thb]
	\begin{center}
		\includegraphics[width=.7\textwidth]{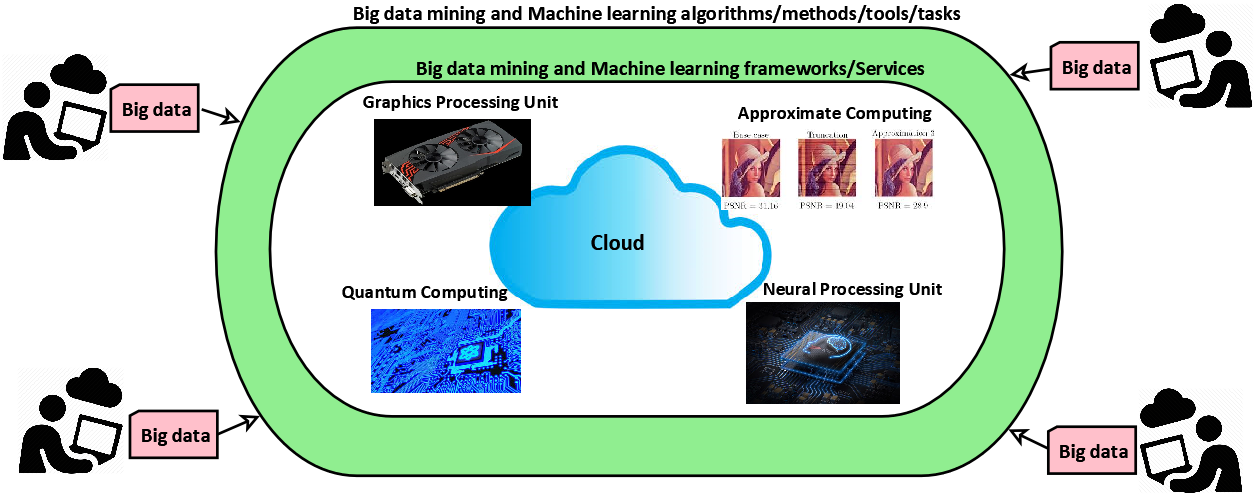}
		\caption{The general architecture using GPUs, QC, AC, and NPUs for achieving acceleration and greenness in big data mining, analytics, and machine learning.}
		\label{fig:general}
	\end{center}
\end{figure}

The figure \ref{fig:general} is a depiction of the cloud based holistic architecture for big data mining and machine learning using the above discussed technologies.
The innermost layer is the cloud computing paradigm in blue color.
The middle layer is the combination of cloud based service and hardware systems consisting of GPUs, NPUs, AC facilities (in the form of hardwares, and softwares) and QC facilities (in white eclipse).
The outermost layer is the combination of big data and machine learning related algorithms and tasks for the user to invoke (in green eclipse).
Then finally comes the cloud service users which have the big data to be mined or analysed or use the data for learning purpose.

The main motivation of this chapter lies in the fact that big data is growing even bigger and bigger. Its not convenient to rely entirely on cloud computing resources.
So, we have to rethink the wheel and reinvent the way big data can be managed and processed for the various applications.
The current literature do not bring the three above mentioned technologies (GPU, AC, and QC) into one paper for discussion and understanding their capabilities in terms of big data mining and machine learning.
We want to put these three mentioned technologies under one umbrella for the researchers to have an easy access and guide of the scenario.
This will help the researchers to delve deeper into these areas and come up with efficient data mining for big data and machine learning tasks in future generation cloud systems.
We have also commented on the future prospects of these technologies and some other new technologies in the section \ref{sec:discussions}.

This chapter is managed into the following sections.
The section \ref{sec:bdm} gives us a short overview about the concept of data mining in big data paradigm and analytics in cloud environments.
The various elements involved in data mining in the big data paradigm has been discussed.
We also discuss, why cloud alone is not sufficient to handle the data outburst of today.
Section \ref{sec:graphical} summarizes the recent works related to the use of GPUs in big data mining and machine learning.
The section \ref{sec:approximate} puts forward the concept of AC and its uses in various machine learning and big data analytics related applications.
Section \ref{sec:quantum} gives some recent works on QC and its applications and implementations in big data and machine learning related algorithms and techniques.
Finally, we comment on the three said technologies (AC, QC, and GPUs) and their impact on big data as a whole in section \ref{sec:discussions}.
We also talk about future research directions in each of the technologies with some new additions of concepts and methods in cloud related environments.
We have also commented on a new technology called Neural acceleration or Neural Processing Units (NPUs). This can be explored to get benefits as per the current literature.
The chapter has been concluded with a short summary in section \ref{sec:conclusion}.
\section{Big Data Mining and Analytics in the Cloud}
\label{sec:bdm}

Data mining has been an important task in many areas of computing for many years. 
Many multi-disciplinary fields are also highly impacted by data mining and analytics. 
Data mining simply is the mining of data or discovery of knowledge from structured, semi-structures unstructured data.
Now, data analytics is a broader term.
It is the process of extraction, cleaning, transforming, modeling and visualizing the data to find meaningful information and further draw inferences and conclusions out of it.
It also involves active machine learning in this process.
We define it as follows:
\begin{equation}
\begin{split}
Data~Analytic~=~Data~Extraction~+~Data~Cleaning~+
~Data~Transforma\\
tion~+~Data~Modeling~+~Data~Visualization~+~Inference~+~Conclusion
\end{split}
\label{eq:da}
\end{equation}

But as the need for more computing power arises, we cannot be sufficient with general computing techniques and methods in our conventional server/workstation setups. 
So, the solution to this problem is the paradigm shift towards cloud computing.
Data mining in the cloud is the concept of performing data mining tasks in a cloud or cloud-like setup.
By cloud-like setup, we mean that we have access to a considerably huge amount of resources in the form of storages, compute nodes, network infrastructure, and other related services. 
And, this is what we call big data mining in the cloud.
The knowledge mined from data is bigger and more complete if the data to be mined is bigger.
\begin{equation}
Big~Data~->~Big~Knowledge~+~Big~Intelligence~+~Big~Cognition
\label{eq:bd}
\end{equation}

The need for big data mining and analytics is evident in many applications.
Mostly, for the applications related to cognition, intelligence, and prediction, data plays a very big role and big data is icing on the cake itself.
The benefits of big data can be realized in the form of big knowledge and big Intelligence.
But the issues of big data are volume, velocity, variety, and veracity.

But is cloud sufficient to handle these features of big data \cite{barua2019comprehensive}?
We have to use some other technologies other than general cloud resources.
We have identified three relevant technologies (GPU, AC, and QC) to achieve efficiency and efficacy in cloud data mining and analytics though we do not claim this is the exhaustive list of the technologies available for this purpose.
The different advantages of using these technologies can be represented in equations \ref{eq:gpu}, \ref{eq:ac}, and \ref{eq:qc}.
Also, some feasible combinations of technologies for future explorations and implementation are shown in equations \ref{eq:acgpu}, \ref{eq:qcgpu}, and \ref{eq:qcac}.
Even we can harness the combination of all these three technologies for big data mining, analytics, and machine learning to achieve greenness, acceleration, and efficiency in a cloud-based setup as shown in equation \ref{eq:acqcgpu}.

\begin{equation}
Acceleration~in~big~data~mining~->~Big~data~in~(GPU + cloud~computing)
\label{eq:gpu}
\end{equation}
\begin{equation}
Greenness~in~big~data~mining~->~Big~data~in~(AC + cloud~computing)
\label{eq:ac}
\end{equation}
\begin{equation}
Efficiency~in~big~data~mining~->~Big~data~in~(QC + cloud~computing)
\label{eq:qc}
\end{equation}

\begin{equation}
\begin{split}
(Acceleration + Greenness)~in~big~data~mining~->~Big~data~in~(GPU + AC\\
+ cloud~computing)
\end{split}
\label{eq:acgpu}
\end{equation}

\begin{equation}
\begin{split}
(Acceleration + Efficiency)~in~big~data~mining~->~Big~data~in~(GPU + QC\\
+ cloud~computing)
\end{split}
\label{eq:qcgpu}
\end{equation}

\begin{equation}
\begin{split}
(Efficiency + Greenness)~in~big~data~mining~->~Big~data~in~(QC + AC\\
+ cloud~computing)
\end{split}
\label{eq:qcac}
\end{equation}

\begin{equation}
\begin{split}
(Acceleration + Efficiency + Greenness)~in~big~data~mining ~->~Big~data~in~\\(GPU + QC + AC + cloud~computing)
\end{split}
\label{eq:acqcgpu}
\end{equation}
\section{Graphical Processing Units (GPUs) and Cloud Big Data Analytics}
\label{sec:graphical}

\begin{figure}[thb]
	\begin{center}
		\includegraphics[width=.5\textwidth]{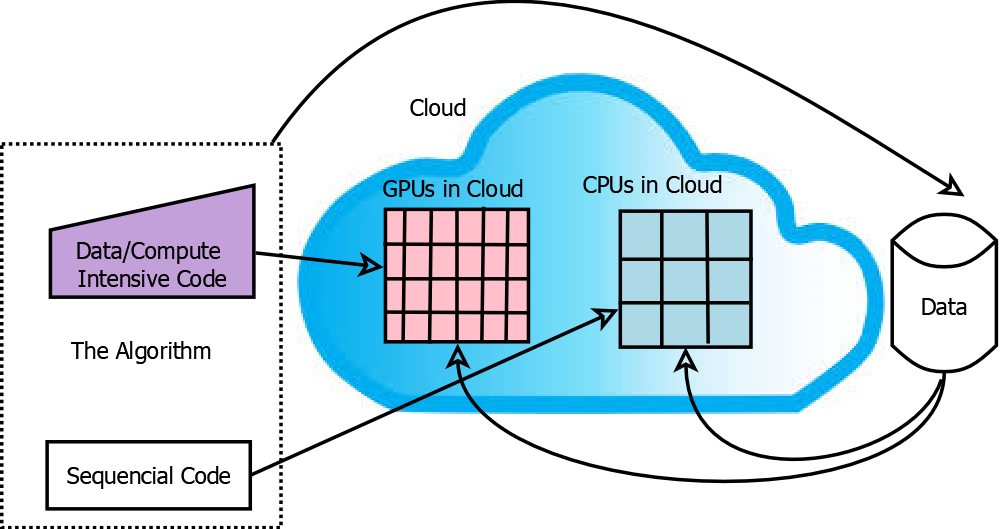}
		\caption{The general architecture for using GPUs in Cloud Big Data Analytics.}
		\label{fig:gpu}
	\end{center}
\end{figure}

GPUs are specifically designed for graphics-related operations. 
These specialized units have many smaller processing units/threads. 
They are best suited for doing parallel processing of complex mathematical operations such as matrix operations. 
Until the recent past, these GPUs have been used for video rendering and multimedia or graphics processing specifically. 
But, people argued that such powerful processing units may be used for other computations in general rather than just using them in graphical utilities such as video rendering, games, and others. 
Hence, comes the concept of General Purpose GPU (GPGPU).
Figure \ref{fig:gpu} shows a typical cloud-based GPU setup to accelerate big data mining and analytics related tasks.
We can see, the compute and Data intensive code parts are delegated to the Multi-core GPUs and the other parts are processed in CPUs in the cloud.
The GPUs help in accelerating the relevant code segments and process data in a much faster manner.

The paper \cite{zhao2014gpu} studies the uses of GPUs in data intensive applications using map-reduce and other graph processing technologies.
The authors report their experiences with developing various platforms for data intensive applications and prototypes for the same purpose. 
Another article \cite{verma2016scaling} is a survey of usage of GPGPUs for cloud frameworks for data intensive computations like big data processing, mining, and analytics.
This paper also suggests CPU-GPU based hybrid techniques for future exploration by the interested researchers.
The authors think that it has a huge prospect in cloud-based big data processing and analytics activities.
Another paper \cite{cano2018survey} has surveyed the use of GPU based systems and computing paradigms on large-scale general data mining or big data analytics tasks.
It discusses the GPU architectures necessary for handling high volume, velocity and variety of data.
Finally, it discusses the limiting causes for proper scalability of such systems in the cloud and some notes on future directions and open research challenges.
The forthcoming paragraphs give some of the major contributions on the usage of GPU, Multi-GPU systems and GPGPU systems in big data mining, analytics, and machine learning.
Other GPU-CPU hybrid techniques used for big data mining in the cloud or related environments have also been mentioned.

In \cite{zhong2013towards}, Zhong et al. put forward a GPU based method, G2 for graph processing in the cloud.
They have implemented three GPU related optimizations:
\begin{itemize}
	\item The first one being some APIs to facilitate the gigantic amount of threads in GPU parallelization.
	\item The second is the introduction of a load balancing method for CPU/GPU architectures using graph partition.
	\item Thirdly, a memory management system is being incorporated transparently in GPU.
\end{itemize}
A task scheduling strategy for high throughput in the form of graph tasks using concurrent kernel executions is also implemented.
Amazon EC2 virtual cluster of eight nodes is being used to perform experimentation.
It shows the efficacy of using GPU acceleration in graph processing tasks in the cloud.

Hai Jiang et al. in \cite{jiang2014accelerating} puts forward a multi-GPU solution to Map-Reduce (MGMR) which can be used for heavy data processing.
This is advantageous over single GPU options too as it eliminates the memory limitations of a single GPU based system.
It also avoids atomic operations for acceleration of the entire process.
The experimental results have shown the efficacy of these techniques using GPUs in handling huge data in the cloud. 
Again, \cite{chen2013pipelined} gives us an advanced version of MGMR, a Multi-GPU pipelined system (PMGMR) to attend the memory limitations of a simple multi-GPU system and high computational demands for big data.
The system has the capacity to use features such as streams and Hyper-Q.
PMGMR is 2.5 times more improved and efficient in terms of performance.
Due to this, its scalability factor increases a lot and user can use it in a simple manner to write map-reduce related codes effectively.
Yet another improvement over the above two GPU based Map-reduce systems is researched in \cite{jiang2015scaling}.
MGMR++ is optimized for big data processing and analytic.
This system even uses hard disks memory when the CPU and GPU memory gets exhausted.
This system also has a 2.5 times improvement in performance compared to MGMR \cite{jiang2015scaling}.

Authors in \cite{abe2018data} propose a very interesting service for the cloud from the perspective of big data and related services.
DaaS (Data as a Service) for analytics of real-time data using GPUs is the main contribution of this paper.
This system is optimized for network data, customer data and user data which are obtained from data centers and cloud-based systems.
The pre-processing module of the DaaS is GPU enabled for fast and efficient processing and accelerating the entire process.
The model is being experimented on huge spatiotemporal data using clustering, self-organizing maps, and neural networks.
The promising results suggest that DaaS using GPU can help in achieving SLA (Service Level Agreement) and QoS (Quality of Service) with great efficiency.

The paper \cite{kurkure2017machine} is focused on GPUs for machine learning (ML) applications in the cloud.
Virtualized GPU techniques have been used as High Performance Computing (HPC) method.
The two main methods referred to are using Hypervisors and vendor-provided virtual GPU technologies.
The paper compares results using virtualized GPUs and also by allowing other applications such as 3D-graphics related tasks so as to utilize the computing power in an efficient manner. 
Some bench-marking ML applications using TensorFlow have been selected for showing scaling between one and multiple GPUs setup.
The paper also compares the performance of the two selected virtualization methods.
In the end, the paper suggests that it is better (in terms of execution time mitigation) to run ML tasks and other typical GPU tasks together in a mixed fashion rather than running them individually.

A very beautiful comparison of non-GPU and GPU enabled big data clustering process is put forward in \cite{adiyoso2018time}.
The paper uses a multi-CPU spark system and a multi-GPU TensorFlow enabled system for unsupervised big data learning.
The later shows a 5-12 times improvement in time over the former.
Jun Wang et al. in \cite{wang2017big} states the issues with the big data pre-processing tasks.
These tasks typically take huge time and computing power in identifying the multi-level hierarchy in big data.
Also, current data have very high dimensionality and it is difficult to scale learning algorithms in such a situation.
So, there is an ardent requirement to device a scalable solution to create a tree structure for big data.
Incremental K-means is used to serve this purpose.
Dimensionality reduction is also incorporated as a part of pre-processing.
The underlying architecture used is CUDA (Compute Unified Device Architecture), so as to facilitate big data requirements.
It has proved efficient with real-world data after visualization using a dendrogram.

The research in \cite{jurczuk2019multi} captures the problem of scalability in Evolutionary Decision Trees (DT). 
It is a hybrid approach of CPU+GPU computing, where the tree structure is searched in a sequential manner in CPU and the fitness is calculated in GPU. 
As a result, the process is accelerated by thousand times using 4 GPUs while using a data-set of about 1 billion objects.
Youcef Djenouri et al. \cite{djenouri2019exploiting} exploits the prospects of GPU cluster computing.
Frequent item-set mining in a single scan has been chosen for the experiment to reduce time complexity.
The authors have proposed a total of three HPC based techniques for the purpose.
The first one uses a GPU based method to efficiently map thread blocks to input blocks.
In the second method, a cluster architecture is used to schedule jobs independently to workers belonging to that cluster.
The third one is a multi-cluster architecture consisting of GPUs to accelerate the frequent item-set mining task. 
Apart from these, several strategies have been incorporated for GPU thread to reduce divergence and load imbalance in clusters.
The third technique has been seen to outperform the first and the second one in speedup parameters, specifically 350 times faster for low minimum support from big data perspective.

The paper \cite{aqib2020memory} puts up an interesting method of road traffic incident prediction using a huge amount of data with deep learning methods on GPU platforms.
The deep learning method uses three different types of data: road traffic-related data, vehicle detector station, and incident data. 
These data are taken from California Department of Transportation (Caltrans) Performance Measurement System (PeMS). 
The research published in \cite{cuzzocrea2019novel} focuses on the utility of big data mining on the Internet of Things (IoT) related applications. 
The authors have used a GPU-aware architecture using Histogram-based segmentation of moving objects. 
The method takes into account the pixel oriented approach pertaining to GPUs (called pixHMOS\_gpu). 
The experimentation has proved the efficacy of this technique bridging the gaps of computational bottleneck often associated with IoT based systems for big data processing and mining.

Patryk Orzechowski et al. \cite{orzechowski2019ebic} gives an efficient bi-clustering method for high volume big data using GPUs. 
Evolutionary search-based bi-clustering (EBIC) has been implemented successfully using multi-GPUs for achieving high scalability. 
The applicability of this method can be found in RNA-sequencing related experiments.
The paper \cite{gonzalez2019accelerating} uses CUDA enabled GPUs for performing complex binary bi-clustering in data mining applications.
The authors present CUBiBit to accelerates the binary bi-clustering tasks by using CPUs and GPUs and CUDA architectures. 
The experimental results depict its amazing speedup of 116 compared to the current method BiBit having 16 CPU cores and three NVIDIA K20 GPUs.
GPU based CUDA processing architectures for spatial data mining is discussed in \cite{oh2019parallel}.
The spatial data here consists of spatial and non-spatial attributes.
So, it is difficult to process and mine such data with general processors or even parallel processing units.
The author has proposed a CUDA based architecture for the same where the experiments have been conducted on TIGER/Line data from US census. 
The results have displayed the superiority of this technique for spatial data mining applications.
\section{Approximate Computing (AC) and Cloud Big Data Analytics}
\label{sec:approximate}

\begin{figure}
	\begin{center}
		\includegraphics[width=.8\textwidth]{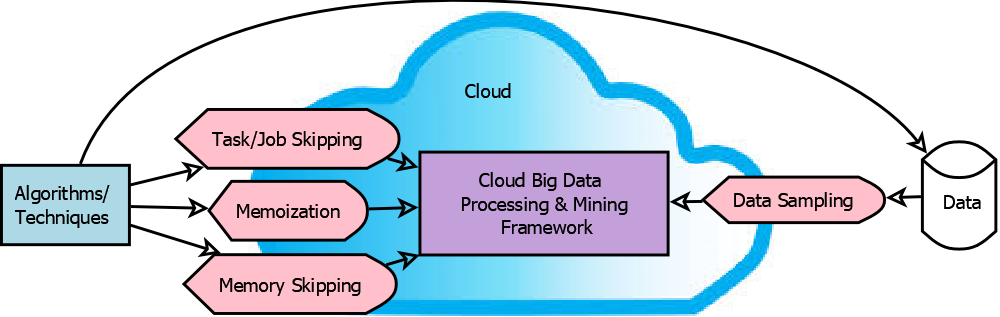}
		\caption{The general architecture for using AC in Cloud Big Data Analytics.}
		\label{fig:AC}
	\end{center}
\end{figure}

Approximate computing or In-exact computing is a technique for trading off result accuracy with speed and energy.
It has been successfully used in many domains ranging from machine learning to financial data analysis.
Big data is a very strong and important target for AC as big data doesn't always require exact analytics output but needs a summary of the output.
Losing a few data items of big data while performing analytics will not impact the final result of mining and analytics. 
A very small change in the data often doesn't have the caliber to change the meaning of the predicted value or shift its importance \cite{nair2014big}.
Some recent articles \cite{nair2014big,barua2019approximate} state the  techniques and applications of AC used in big data and machine learning perspective in the cloud and related distributed platforms.
The remainder of this section discusses the usage and applications of AC in big data mining and machine learning in cloud environments.
Figure \ref{fig:AC} gives an idea of the architecture for using AC techniques atop cloud data mining or analytics frameworks.
The major techniques are task/job skipping, memoization and memory skipping. 
Data sampling can also be applied to cut down unnecessary data and use only a small subset of it for further processing without much change in the mining output.

Shuai Ma et al. \cite{ma2019approximate} puts forward the case of big data analytics and its applications in today's world.
Big data analytics need huge computational and storage power.
To harness the optimal solutions from big data it is necessary that a huge amount of computing is devoted.
But in that case, the efficiency of the systems may degrade as the data increases.
So, if the optimal solution is replaced with some "good enough" solution which is acceptable than it is a desirable situation for all.
Techniques such as approximate query processing and data approximations are being employed in big data analytics.
Basically, the query approximation technique has been used in pattern matching in graphs, compression of trajectory and computations of dense sub-graph.
Moreover, the data compression technique has been used successfully in shortest path computation, network anomaly detection and link prediction in graphs for social media analytics.

Barua et al. \cite{barua2018green} have proposed a method using loop perforation technique of AC in association rule mining.
The new approach cuts down the execution time of the mining task with a loss of accuracy in finding the rules which is acceptable in case of big data volumes.
This method can be implemented in parallel systems such as a cloud to attain more efficiency and performance improvement.
In \cite{ahmadvand2019gapprox}, authors address the issue of dependency on big data mining results on data variety and its impact on using AC. 
Using AC when the data is big but variety is low is a trivial task as per the techniques available nowadays such as sampling, memoization and task skipping, etc.
But the same thing, if used in high variety big data sets, can be disastrous as skipping data from a certain part of such big data may skip a whole lot of a variety from it causing unacceptable results.
So, the paper \cite{ahmadvand2019gapprox} has proposed Gapprox, which is a data variety aware system for approximate computing.
The method used is a sampling technique after clustering the data into relevant buckets using intra/inter-cluster distance.
The size of the blocks and samples are optimized for Quality of Result and acceptable confidence and error bound.
The experimental results show that it outperforms ApproxHadoop \cite{goiri2015approxhadoop} by 17x speedup and Sappox \cite{zhang2016sapprox} by 8x with 5\% error tolerance by a user with 95\% confidence.

The authors of \cite{goiri2015approxhadoop} have put forward an approximate version of the map-reduce paradigm (ApproxHadoop) using task skipping or dropping and data sampling methods.
The error bounds for various map-reduce programs have been theoretically determined using statistical formulas.
The major applications targeted are data analytics, scientific computation, and machine learning.
The speedup achieved is about 32x if the user is ready to tolerate an error of 1\% having 95\% confidence.

Xuhong Zhang et al. \cite{zhang2016sapprox} have put forward an approximation method for arbitrarily chosen sub-data-sets from large data-sets.
Generally, if the sub-data-sets are uniformly distributed, sampling works pretty well.
But for unevenly distributed sub-data-sets the sampling efficiency is very low resulting in poor accuracy in the estimation.
For this reason, a distribution-aware method is required.
The logical partition of a data-set having sub-data-sets is examined for occurrences.
This information is used for online sampling.
Hadoop is used to implement this method called Sapprox.
The results show that it is 20x times faster than its precise version.

A novel approach to approximate computing using a neural network (NN) architecture is presented in \cite{peng2018axnet}.
The said NN architecture is a combination of two NNs, one being the approximator and another is the predictor.
The approximator gives us the approximate version of the results.
The predictor does the quality check by determining whether the input data is eligible for approximation for a given output accuracy requirement for any application.
It is not easy to create such a combined NN having two different NNs as there will be coordination issues and different optimization objectives.
So, AXNet is a combination of approximator NN and predictor NN to create a new holistic NN having end-to-end trainable capabilities.
Multi-task learning is being used in AXNet to find a better and higher number of approximable candidate samples.
The error due to approximation is minimized and training cost is mitigated too.
The experimentation with this novel NN architecture shows its efficacy in reducing training time and determining approximable samples.

The usage of Spiking Neural Networks (SNNs) for data-intensive applications such as analytics and vision is discussed in \cite{sen2017approximate}.
It is difficult to find adequate compute and storage power for large scale SNNs.
Hence, the paper proposes AxSNN atop the already proposed parallel version of the same.
It is being used to improve the computational efficiency of SNNs in general.
SNNs work with inputs and outputs of neurons which are generally represented as time series of spikes.
The internal states of a neuron are updated by spikes at the output of a neuron which is connected to it.
AxSNN leverages the notion of approximate computing by skipping neuron updates triggered by spikes considering that it has minimum impact on the quality of output.
This is done to improve the computing energy and memory efficiency.
Parameters that are considered for approximating neurons are average spiking rates, current internal states of neurons and the weights of synaptic connections.
Approximate computing on a neuron is attained by making it sensitive to a subset of its inputs and sending spikes to a subset of its outputs.
The overall system is monitored and approximation modes are updated such as the energy savings are optimized and quality loss is minimum.
It has been tested in the form of hardware and software implementations both.
These have been tested in many image recognition applications and it achieves 1.4-5.5x reduction in operations.
It is equivalent to a 1.2-3.7x reduction in energy approximately in hardware and software implementations without any loss of output.

The paper \cite{hanif2018error} puts forward a method for approximating Convolution Neural Networks (CNNs) for image processing, big data analytics, computer vision, mining, AI and ML related applications.
The paper proposes a systematic method for checking the error tolerance characteristics of a deep CNN and determine the parameters from the set of parameters which can be targeted to improve speed of the network in the inference stage.
The idea is to cut down the filters as per their significance in a convolution layer.
This enables a trade-off between output quality and speedup.

Authors in \cite{sen2018approximate} give us a methodology for introducing efficient approximate computing in Recurrent Neural Networks (RNNs) - Long Short Term Memory (LSTM).
LSTMs are generally used for text generation, speech recognition, and related application areas.
Generally, such applications in a big data perspective can be computation-intensive in nature.
It is difficult to address such issues with distributed, parallel cloud computing setups.
To overcome these constraints, the authors have proposed AxLSTM (Approximate LSTM).
AxLSTM is being used in sequence-to-sequence learning using TensorFlow deep learning framework.
It achieves a speedup of about 1.31x maximum with no loss of accuracy of output and 1.37x speedup when acceptable reductions in output quality are permitted.

Do Le Quoc et al. \cite{quoc2017privacy} puts forward a data analytics system for stream processing with privacy preservation.
PRIVAPPROX is a system which gives us a low latency stream analytics with proper privacy preservation for users.
There are many features of this framework, but the most important one being the ability to determine the optimum trade-offs between output quality with minimizing error and the execution cost.
It is also reported to have the ability to do real-time stream analytic using distributed system setups.
The main idea of this framework is the combination of sampling and randomized response.
Due to this combination of two unique features, it is a system with high privacy for users and high performance in terms of execution time and energy savings.

The article \cite{quoc2017streamapprox} aims at presenting a stream processing system with less computation and response time achieved by AC. 
The authors have also taken care of error bounds so that the system does not fail with excessive approximation.
The algorithm is designed in such a way that it can be used with Apache Spark batched Streaming and pipeline based Apache Flink as well.
StreamApprox is the full prototype of the described system atop Apache Spark Streaming and Apache Flink.
The experimentation on real-world case studies shows that it has a speedup of at least 1.2x and a maximum of 3x compared to the classical counterparts.

The paper \cite{wen2018approxiot} targets approximate computing for stream analytics in real-time for IoT devices.
The paper proposes APPROXIOT which employs an online hierarchical classified reservoir sampling technique that produces approximate output with defined error bounds.
Edge computing resources are being used to realize this technique.
Apache Kafka is being used as the underlying framework to implement this method.
A set of real-world case studies have been used for evaluation, which shows that it attains a speedup of 1.3x-9.9x when the sampling fraction is being changed from 80\% to 10\%.

The work in \cite{krishnan2016incapprox} gives a data analytics method for incremental approximate computing.
Incremental computing is based on memoization of intermediate results in a multi-step process where the intermediate output is memorized for sub-computations.
AC means skipping jobs or tasks or sampling huge data-sets into sub-data-sets for efficient use.
The paper explores these two methods by designing a sampling algorithm that selects samples on the basis of memoized data from previous runs.
It is based on a self-adjusting computation that produces output with error bounds.
It is termed as IncApprox and is based on the Apache Spark Streaming framework.
The experimentation on real-world case-studies confirms the system to have used incremental and approximate computing delivering high-efficiency benefits.
\section{Quantum Computing (QC) and Cloud Big Data Analytics}
\label{sec:quantum}

The concept of Quantum computing \cite{rieffel2000introduction, denchev2008distributed, lo1998introduction} emerged in 1980.
The first quantum mechanical model was described by Paul Benioff, showing that theoretically there is the full possibility of such a computer.
The Quantum computing paradigm is not restricted to the two conventional states of a classical computer, i.e., '0' and '1'.
A quantum computing setup can have a superposition of such states which are called qubits. 
Quantum computing has the ability to solve some computing-intensive problems in a much faster manner than their classical counterparts.
The paper \cite{shaikh2016quantum} reviews the utility of quantum computing in big data processing and machine learning related applications and its current scenario in the research community.
The paper discusses Quantum Artificial Neural Networks which we may call as QANN.
It can speed-up the learning process of any conventional neural network by many folds.
The use of quantum computing in supervised and unsupervised learning is being discussed. 
Big data analytics can be highly improved in terms of speed-up with the help of quantum computing.
The paper also discusses the challenges and future prospects of quantum computing in machine learning and related fields.

\begin{figure}
	\begin{center}
		\includegraphics[width=.7\textwidth]{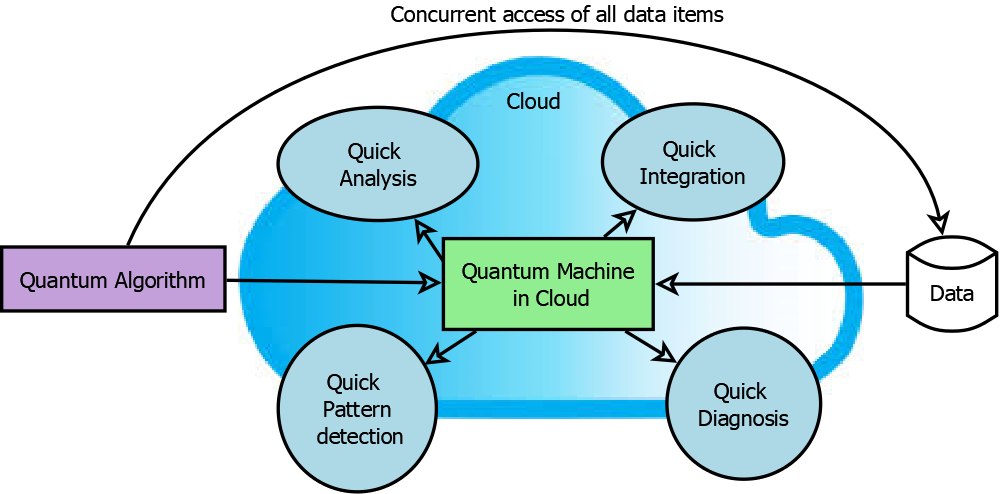}
		\caption{The general architecture of using QC in Cloud Big Data Analytics.}
		\label{fig:QC}
	\end{center}
\end{figure}

The paper \cite{schuld2015introduction} surveys the idea of implementing classical machine learning algorithms or their specific computation hungry sub-parts in quantum setups.
Most of the stochastic methods can be designed in a way that it can be implemented in quantum systems.
The paper gives a very clear picture of quantum machine learning and puts forward the high-level descriptions of the existing methods and techniques along with the low-level technicalities.
The paper also comments on the future and research prospects of quantum learning theory.
Figure \ref{fig:QC} conceptualizes the idea of QC accelerated mining and analytics in the cloud.
The basic idea is the use of quantum algorithms on quantum machines in cloud setup which can concurrently access the data in hand for processing, analysis, integration and pattern detection.

The paper \cite{biamonte2017quantum} is a review article that speaks about the promises of quantum techniques to solve machine learning problems and data pattern mining applications more efficiently than classical systems.
This is due to the fact that the quantum computing paradigm is ought to work in such a way that it can outperform classical systems in certain tasks specifically.
The new field, that has been talked about in this article speaks about Quantum Machine Learning (QML) that is dedicated towards implementing quantum algorithms and techniques and softwares which can model machine learning systems and programs.
But the real challenge in such a scenario is the hardware that can support quantum systems and software.

An idea about how quantum computing can be utilized for big data-related technologies is shown in \cite{wang2015big}.
Big data has seen a tremendous explosion in recent years and it is becoming bigger and bigger with each passing day.
With such a situation it is not possible to rely simply on cloud resources such as parallelism, distribution, huge compute power and storage alone for the sake of better and efficient analytics.
The main concept of quantum computing allows it to perform quantum parallelism and can be very fast as compared to classical systems even with cloud support.
The paper discusses the Grover search algorithm.
Quantum machine learning can be used to implement big data analytics through the eye of quantum computation. 
The paper discusses the main applications of quantum computation in data mining and related tasks.

Patrick Rebentrost et al. \cite{rebentrost2014quantum} discuss the implementation of a supervised learning technique in quantum computing architecture.
Support Vector Machine (SVM) has been chosen for this experiment and good results have been achieved.
Generally, the SVM takes polynomial time complexity in a classical computer.
It has been implemented in logarithmic time complexity in quantum setup displaying an exponential speedup in the process.
This big data related SVM technique is based on the non-sparse matrix exponentiation technique.
Another research work \cite{havlivcek2019supervised} discusses the limitations of a machine learning problem considering SVM specifically.
The feature space determination and its size are the real problems in machine learning and these need adequate attention.
The quantum algorithms have quantum state spaces using entanglement and interference which are really large if compared to classical counterparts.
The paper proposes algorithms on the basis of quantum state spaces which can be used as feature space for ML problems.
The quantum variational classifier has a variational quantum circuit and the quantum kernel estimator can optimize a classical SVM by estimating the kernel function on the quantum setup.

In \cite{schuld2014quantum} the prospects of quantum computing in pattern classification of big data have been discussed.
The conventional machine learning algorithms can be enhanced in terms of performance using quantum information theories.
The paper introduces the quantum pattern classification algorithm and its application in handwritten digit recognition from the MNIST data-sets.
Article \cite{ruan2017quantum} gives us a quantum version of K-nearest neighbors (QKNN) using a metric of Hamming distance.
A quantum circuit is being created and implemented to calculate Hamming distances in testing samples and the feature vectors in the training set.
The QKNN method achieves good performance in terms of complexity and also shows good classification accuracy.
This method has proven itself to be better than many existing methods.

The paper \cite{sheng2017distributed} discusses the idea of offloading a quantum machine learning task from a classical computer to a remote quantum computer with proper data privacy preservation methods implemented atop it. 
Distributed secure quantum machine learning (DSQML) is a method which is proposed to do the said job.
It not only offloads ML tasks but also accesses remote data in the database server.
A robust protocol is designed to prevent any eavesdropping or any hindrance in the learning process.
The protocol is made for the classification of high dimensional vectors and can be a good candidate to be used in big data applications in the future.

Ashish Kapoor et al. \cite{kapoor2016quantum} state that the use of quantum computation methods can bring about a drastic improvement in the computational complexity of perceptron learning.
The paper gives two algorithms, one uses quantum information processing and another shows that the mistake bound of classical computing can be improved by quantum computing.
Zhaokai Li et al. \cite{li2015experimental} show a quantum parallelism based machine learning algorithm to implement handwriting recognition.
The quantum machine which is used for this experiment has a 4-qubit NMR test bench.
In today's age of artificial intelligence and big data, such a quantum-based method could be very lucrative in the long run.

The paper \cite{hu2019quantum} gives us a quantum computing approach for image processing and computer vision-based applications like object detection.
Object detection is one of the most important applications in many areas such as surveillance, robotics and many more.
The research in this paper presents an automated object detection method using quantum techniques.
The experimentation results have proved the supremacy of such a quantum-based algorithm in object detection accuracy which minimizes measurement errors in general.

Matthew C. Johnson et al. \cite{johnson2019integration} have proposed the marriage of quantum computing devices and distributed computing paradigms, which can be thought of as a service in the cloud as well.
The system has APIs and data models for quantum computing.
Some softwares are also in place for creating these quantum data models and compute them to get the relevant results from the quantum devices.
This way the distributed computing paradigms can be made more effective for high performance by integrating quantum device into their setups.

\section{Discussions, Prospects and Future Trends}\label{sec:discussions}
The chapter has categorized literature into three primary buckets for the three primary game-changing technologies.
However, we want to have a better insight into the technologies and trends from the lens of big data, cloud data mining, and machine learning.
We also commented on the future of the three technologies along with NPUs which have not been discussed extensively in the chapter like the other three.

The table \ref{tab:object} gives us a first-hand list of all the literatures cited in this chapter on the basis of the discussed technologies for acceleration (GPU and QC) and greenness (AC) and application areas in cloud and related environments.
It shows that GPU and AC have dominated the big data mining and analytic area in recent times as compared to machine learning.
But, QC has more impact on machine learning than big data mining.
So, a combination of these technologies can be thought of as new avenues for researchers to explore and architect frameworks and platforms for big data and machine learning.

\begin{table}
	\centering
	\caption{A classification based on different technologies used for big data mining and machine learning in cloud}
	\centering
	\begin{tabular}{p{5cm}|p{2.5cm}|p{5.5cm}|}
		\cline{1-3}
		\multicolumn{1}{|c|}{Technology used} & Application area & References \\ \cline{1-3}
		\multicolumn{1}{|c}{\multirow{2}{*}{GPU}} & \multicolumn{1}{ |c| }{Big data mining}  & \cite{zhao2014gpu}, \cite{verma2016scaling}, \cite{cano2018survey}, \cite{bhargavi2017accelerating}, \cite{zhong2013towards}, \cite{jiang2014accelerating}, \cite{chen2013pipelined}, \cite{jiang2015scaling}, \cite{abe2018data}, \cite{adiyoso2018time}, \cite{wang2017big}, \cite{jurczuk2019multi}, \cite{djenouri2019exploiting}, \cite{cuzzocrea2019novel}, \cite{orzechowski2019ebic}, \cite{oh2019parallel}, \cite{gonzalez2019accelerating} \\ \cline{2-3}
		\multicolumn{1}{|c}{}                     & \multicolumn{1}{ |c| }{Machine learning} & \cite{bhargavi2017accelerating}, \cite{kurkure2017machine}, \cite{adiyoso2018time}, \cite{aqib2020memory} \\ \cline{1-3}
		\multicolumn{1}{|c}{\multirow{2}{*}{AC}}  & \multicolumn{1}{ |c| }{Big data mining}  & \cite{nair2014big}, \cite{barua2019approximate}, \cite{quoc2017streamapprox}, \cite{barua2018green}, \cite{goiri2015approxhadoop}, \cite{zhang2016sapprox}, \cite{ahmadvand2019gapprox}, \cite{krishnan2016incapprox}, \cite{quoc2017privacy}, \cite{wen2018approxiot}, \cite{ma2019approximate} \\ \cline{2-3}
		\multicolumn{1}{|c}{}                     & \multicolumn{1}{ |c| }{Machine learning} & \cite{sen2017approximate}, \cite{hanif2018error}, \cite{sen2018approximate}, \cite{peng2018axnet} \\ \cline{1-3}
		\multicolumn{1}{|c}{\multirow{2}{*}{QC}}  & \multicolumn{1}{ |c| }{Big data mining}  & \cite{wittek2014quantum}, \cite{shaikh2016quantum}, \cite{wang2015big}, \cite{ruan2017quantum}, \cite{johnson2019integration} \\ \cline{2-3}
		\multicolumn{1}{|c}{}                     & \multicolumn{1}{ |c| }{Machine learning} & \cite{wittek2014quantum}, \cite{schuld2015introduction}, \cite{biamonte2017quantum}, \cite{wang2015big}, \cite{rebentrost2014quantum}, \cite{schuld2014quantum}, \cite{sheng2017distributed}, \cite{havlivcek2019supervised}, \cite{kapoor2016quantum}, \cite{li2015experimental}, \cite{ruan2017quantum}, \cite{hu2019quantum}, \cite{johnson2019integration} \\ \cline{1-3}
	\end{tabular}
	\label{tab:object}
\end{table}

\begin{description}
	\item[Approximate Computing:]
	AC has been instrumental in reducing time complexity for many cloud and non-cloud operations specifically for big data and machine learning applications.
	The different neural network architectures are the prime targets of AC in machine learning as seen in the literature.
	The targeted algorithms and tasks are stream processing, association rule mining, map-reduce paradigm, Hadoop ecosystem, spark framework, streaming data analytic, spiking neural networks, convolution neural networks, Kafka framework, real-time analytic, recurrent neural networks, and long short term memory.
	The common thing seen in this technology is that it can considerably accelerate the analytics, mining or learning process with energy savings to a great extent. 
	The error bound is also taken care of by various statistical models and allowable to a safe extent. 
	In the future, researches can be focused on implementing approximate computing services in cloud setups, AxCaaS- Approximate Computing as a Service \cite{barua2019approximate}.
	One more area to give focus can be from big data and machine learning perspective, by creating approximate computing facilitated big data analytic platforms or services in cloud.

	\item[Accelerated GPU:] GPUs have been a powerful hardware tool for acceleration in any cloud-based cluster systems or even in a standalone system.
	Generally they are used for multimedia computing and video/image processing applications. 
	But, researchers argued that they can be of great help in general-purpose computing too. 
	Hence, we have a different class of GPUs called GPGPUs or General Purpose GPUs. 
	The different services and algorithms targeted by GPUs are graph processing and mining, large scale data mining, map-reduce paradigm, GPU based Data as a service (DaaS) in the cloud, real-time data analytics, neural networks, self-organizing maps, clustering, machine learning in 3D graphics, spark framework, TensorFlow system, unsupervised learning, k-means, decision trees, deep learning, data fusion, bi-clustering, and bio-informatics.
	The surveyed literature in this chapter tells that graph processing and mining is the most suitable candidate for implementing GPU based acceleration.
	The main three concerns of big data analytics and related technologies are big volume, big velocity, and big variety and GPUs are capable of taking care of all these three elements of big data efficiently. 
	The main advantage of GPUs can be stated as the achievement of service level agreements and quality of service in cloud-based systems.
	Here big data is being harnessed as workflows by various users.
	So, we want to suggest GPU as a service specifically for big data scenarios (BD-GPUaaS) and machine learning in GPU as a service (ML-GPUaaS) in the cloud for advanced machine learning techniques such as deep learning.

	\item[Quantum Computing:] QC is in the focus light for the last few years in many domains of computing and the major reason is its speed of computing some of the conventional things in an un-imaginable manner compared to classical counterparts. 
	Quantum computing has been proposed by many researchers for big data and machine learning related computations. 
	Quantum computing has been widely used to implement machine learning techniques in general.
	The theory of quantum learning and information processing has been well crafted to suit the requirements of these applications.
	The various targeted areas are artificial neural networks, support vector machine, pattern classification, k-nearest neighbors, object detection and recognition.
	Some literatures also talk about quantum computing in a distributed environment such as cloud.
	There they can offload quantum computing-related tasks to quantum computers and other tasks can be done in the in-house system itself. 
	We want to suggest quantum computing as a service for big data and machine learning for the future.
	QCaaS has been mentioned in the paper \cite{barua2019comprehensive} and we hope that it can be a good platform for the researches to conduct fruitful researches on big data and machine learning. 
	We can also think about BD-QCaaS and ML-QCaaS for big data and machine learning in the future to achieve acceleration and performance improvement in the future.

	\item[Neural Processing Unit:] 
	We would also like to give a possible future thought on a new technology that can be used in this respect specifically in machine learning.
	The rise of neural and AI accelerators or Neural processing units (NPUs) are in the process. 
	The neural accelerators \cite{nair2014big,esmaeilzadeh2012neural, wiki:2020:AI} are best suited for neural network computations for machine vision, big data processing and learning from a huge amount of data. 
	The application areas can be robotics, IoT and data-intensive computations at the edge/fog devices. 
	So, we strongly recommend this technology to be actively researched and implemented in cloud-based systems for big data mining and machine learning. 
	The neural accelerators for AI applications are inherently built to approximate the regions of code by changing them to neural models.
	So, indirectly these accelerators are using the AC paradigm and researchers who are working with AC currently can also look into this new area of research and development from big data and machine learning perspective.
\end{description}
\section{Conclusions}
\label{sec:conclusion}

This chapter provides a very concise and lucid depiction of some of the recent technologies which can enhance the cloud or related high-performance computing models.
This will support the efficiency issue of big data mining and analytics or machine learning-related tasks. 
The chapter gives a generic overview of data mining from a big data perspective. 
We have defined big data as an entity consisting of 4Vs (volume, velocity, variety, and veracity) but on a larger scale.
The Cloud computing paradigm is being used successfully for the last decade to facilitate big data analytic and mining, to bring efficiency and performance in the entire analytic process. 
Existing surveys and researches have conveyed that having cloud resources at our disposal does not help in achieving the maximum speedup and complexity benefits in terms of big data.
We have identified three future technologies to solve this situation in a cloud platform.
Two major kinds of performance aware technologies are identified such as Greenness and Acceleration in computation.
Under the umbrella of green computing, we chose approximate computing (AC) and within acceleration, we have chosen GPU based acceleration and Quantum computing (QC) based performance acceleration. 
We have also shown that Neural Processing Units (NPUs) can also prove beneficial for big data and machine learning process acceleration for future generation cloud systems.

\bibliographystyle{plain}
\bibliography{reference} 
\end{document}